\documentclass{article}

\usepackage{arxiv}

\usepackage[utf8]{inputenc} 
\usepackage[T1]{fontenc}    
\usepackage{hyperref}       
\usepackage{url}            
\usepackage{booktabs}       
\usepackage{amsfonts}       
\usepackage{amssymb}        
\usepackage{amsmath}        
\usepackage{nicefrac}       
\usepackage{microtype}      
\usepackage{lipsum}		
\usepackage{graphicx}
\usepackage[numbers]{natbib}
\usepackage{doi}
\usepackage{wrapfig}

\title{Clifford algebra in R}

\author{ \href{https://orcid.org/0000-0001-5982-0415}{\includegraphics[width=0.03\textwidth]{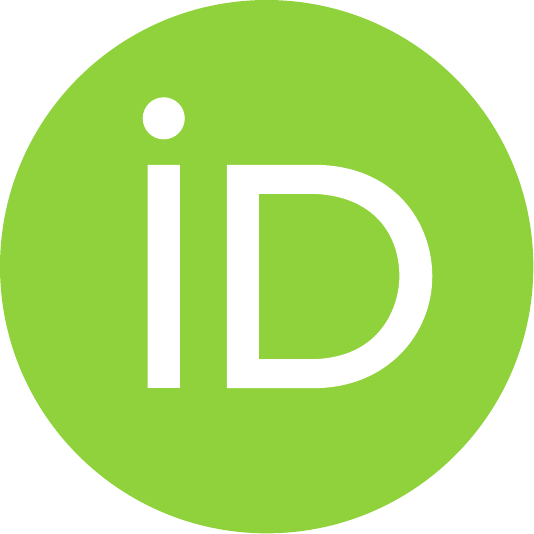}\hspace{1mm}Robin K. S.~Hankin}\thanks{\href{https://academics.aut.ac.nz/robin.hankin}{work};  
\href{https://www.youtube.com/watch?v=JzCX3FqDIOc&list=PL9_n3Tqzq9iWtgD8POJFdnVUCZ_zw6OiB&ab_channel=TrinTragulaGeneralRelativity}{play}} \\
 Auckland University of Technology\\
	\texttt{hankin.robin@gmail.com} \\
}

\renewcommand{\shorttitle}{The \textit{clifford} package}

\hypersetup{
pdftitle={Clifford algebra in R},
pdfsubject={q-bio.NC, q-bio.QM},
pdfauthor={Robin K. S.~Hankin},
pdfkeywords={Clifford algebra}
}

\usepackage{Sweave}
\begin{document}
\maketitle

\begin{abstract}

Here I present the {\tt clifford} package for working with Clifford
algebras in the R programming language.  Algebras of arbitrary
dimension and signature can be manipulated, and a range of different
multiplication operators is provided.  The algebra is described and
package idiom is given; it obeys {\tt disordR} discipline.  The
package is available on CRAN at
\url{https://CRAN.R-project.org/package=clifford}.

\end{abstract}


\section{Introduction}

\setlength{\intextsep}{0pt}
\begin{wrapfigure}{r}{0.2\textwidth}
  \begin{center}
\includegraphics[width=1in]{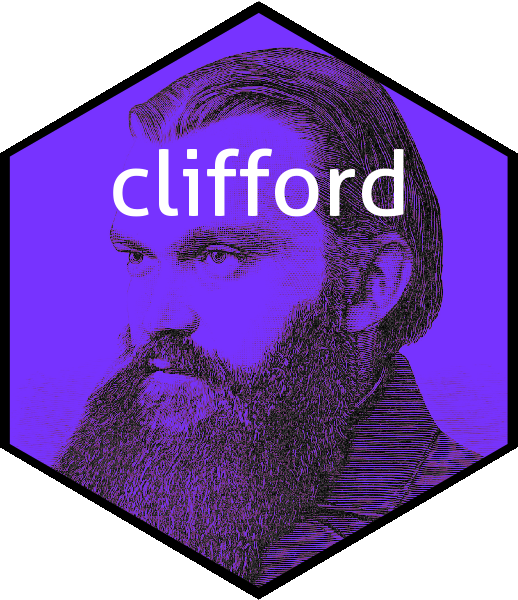}
  \end{center}
\end{wrapfigure}
Clifford algebras are interesting and instructive mathematical
objects.  The class has a rich structure that has varied applications
to physics.  Computational support for working with the Clifford
algebras is part of a number of algebra systems including
Sage~\citep{sagemath2019} and \textit{sympy}~\citep{sympy2017}.  Here
I introduce the {\tt clifford} package, written in the R computing
language~\citep{rcore2022}, which furnishes functionality for working
with Clifford algebras.  Notation follows Snygg~\cite{snygg2010}.

\newcommand{\ei}[1]{\ensuremath{{\bf e}_{#1}}}
\newcommand{\bx}{\ensuremath{{\bf x}}}
\newcommand{\by}{\ensuremath{{\bf y}}}

\subsection{Motivating examples in low-dimensional space}

Following Snygg~\cite{snygg2010}, we consider a vector space of
dimension 3, and given a basis $\ei{1},\ei{2},\ei{3}$, we can consider
linear combinations such as

\begin{eqnarray}
\bx = x^1\ei{1} + x^2\ei{2} + x^3\ei{3}\nonumber\\
\by = y^1\ei{1} + y^2\ei{2} + y^3\ei{3}.
\end{eqnarray} 

where $x^i,y^i$ are real numbers.  A Clifford algebra includes a
formal product on such sums, defined using the relations

\begin{eqnarray}\label{square}
\left(\ei{1}\right)^2=
\left(\ei{2}\right)^2=
\left(\ei{3}\right)^2=1\\
\ei{2}\ei{3} + \ei{3}\ei{2} = \label{sumprod}
\ei{1}\ei{3} + \ei{3}\ei{1} = 
\ei{2}\ei{1} + \ei{1}\ei{2} = 0
\end{eqnarray}

This gives:
  
\begin{eqnarray}
\bx\by &=&
\left(x^1\ei{1} + x^2\ei{2} + x^3\ei{3}\right)
\left(y^1\ei{1} + y^2\ei{2} + y^3\ei{3}\right)\nonumber\\
&=& \left(x^1y^1+x^2y^2+x^3y^3\right) +\nonumber\\&&
       \left(x^2y^3-x^3y^2\right)\ei{2}\ei{3} + 
       \left(x^3y^1-x^1y^3\right)\ei{1}\ei{3} + 
       \left(x^1y^2-x^2y^1\right)\ei{1}\ei{2}
\end{eqnarray}
  
Multiplication is associative by design.  Snygg \cite{snygg2010} goes
on to consider the algebra spanned by products of
$\ei{1},\ei{2},\ei{3}$ and shows that this is an eight dimensional
space spanned by

\begin{equation}
\left\{
1,\ei{1},\ei{2},\ei{3},\ei{12},\ei{31},\ei{12},\ei{123}
\right\}
\end{equation}

where $\ei{12}=\ei{1}\ei{2}$ and so on.  Thus a general element of
this space would be

\begin{equation}
a^0+
a^1\ei{1} + a^2\ei{2} + a^3\ei{3} +
a^{12}\ei{12} + a^{31}\ei{31} + a^{23}\ei{23} +
a^{123}\ei{123}
\end{equation}

(here the $a$'s are real).  That the space is closed under
multiplication follows from equations~\ref{square} and~\ref{sumprod};
thus, for example,

\begin{equation}
  \ei{1}\ei{3}\ei{1}\ei{2}=
 -\ei{1}\ei{1}\ei{3}\ei{2}=
 -\ei{3}\ei{2}=
  \ei{2}\ei{3}=\ei{23}.
  \end{equation}

(observe how associativity is assumed).

\subsection{Generalization to arbitrary dimensions}

Generalization to higher dimensional vector spaces is
easy~\cite{hestenes1987}.  Suppose we consider a $n$-dimensional
vector space spanned by $\ei{1},\ldots,\ei{n}$.  Then an arbitrary
vector in this space will be $a^1\ei{1}+\cdots+a^n\ei{n}$.  The
associated Clifford algebra will be of dimension $2^n$, spanned by
elements like $\ei{1}\ei{3}\ei{5}=\ei{135}$ and
$\ei{1}\ei{2}\ei{3}\ei{5}=\ei{1235}$.  The defining relations would be

\begin{equation}\label{posdefcliff}
\ei{i}\ei{j}+\ei{j}\ei{i}=2n_{ij}
\end{equation}
where  
\begin{equation}\label{posdefcliff2}
  n_{ij} = \begin{cases}
    1, & i=j\\
    0 &i\neq j
  \end{cases}
\end{equation}

\subsection{Clifford algebra in a pseudo-Euclidean space}

Equations~\ref{posdefcliff} and~\ref{posdefcliff2} defined a
positive-definite inner product on the vector space spanned
by~$\ei{1},\ei{2},\ei{3}$.  This is readily generalized to allow a
more general inner product.  Conventionally we define

\begin{equation}\label{gencliff1}
\ei{i}\ei{j}+\ei{j}\ei{i}=2n_{ij}
\end{equation}
where  
\begin{equation}\label{gencliff2}
  n_{ij} = \begin{cases}
    1, & i=j=1,\ldots,p\\
    -1, & i=j=p+1,\ldots,n\\
    0, &i\neq j
  \end{cases}
\end{equation}

for $1\leqslant p\leqslant n$; usually we also specify $p+q=n$ and
write $\mathbb{R}_{p,q}$ for the $p+q$-dimensional vector space with
inner product given by equation~\ref{gencliff1}.  The Clifford algebra
${\mathcal C}_{p,q}$ (other notations include $Cl(p,q)$) is then the
algebra formed by $\mathbb{R}_{p,q}$ together with formal products of
basis vectors.

Note carefully that the diagonal matrix of the inner product specified
above conventionally has the the positive elements first, followed by
the negative elements.  But in relativity, the metric tensor $\eta$ is
usually written with the negative elements first followed by the
positive elements;

\begin{equation}\eta=
  \begin{bmatrix}
    -1&0&0&0\\
    0&1&0&0\\
    0&0&1&0\\
    0&0&0&1\\
  \end{bmatrix}
\end{equation}

\subsection{Wedge product of the exterior algebra is a special case
  of the geometric product}

If we specify that the quadratic form is identically zero then
equation \ref{gencliff1} becomes

\begin{equation}\label{specwedge}
\ei{i}\ei{j}+\ei{j}\ei{i}=0,\qquad 1\leqslant i,j\leqslant p
\end{equation}

which implies that $\ei{i}\ei{i}=0$.  Geometric products become wedge
products (although linearity means that we may add terms of different
grades, unlike conventional Grassman algebra).

\section{Computational implementation and {\tt disordR} discipline}

The package represents basis blades using dynamic bitset objects from
the {\tt boost} library.  A bitset emulates an array of Boolean
elements, but is optimized for space allocation and
access/modification times.  The set bits specify the basis blades
present in a term; using bitsets allows products to use fast Boolean
operators.  An object such as $\ei{2}\ei{5}\ei{6}$ [or $\ei{256}$]
will be a bitset with bits 2, 5, and 6 set [note the off-by-one
  issue].  Dynamic objects are needed here as the number of bits in
the set is specified at runtime.
A {\tt clifford} object is an element of a Clifford algebra; this is
set of basis blades, each with a real coefficient.  The {\tt stl} map
class~\cite{musser2009} is used:

\begin{verbatim}
typedef boost::dynamic_bitset<> blade;
typedef map<blade, long double> clifford;
\end{verbatim}

A ``map'' is a sorted associative container that contains key-value
pairs with unique keys.  It is used because search and insertion
operations have logarithmic complexity.  A {\tt clifford} object thus
maps dynamic bitsets (basis blades) to long doubles in the same manner
as the {\tt spray} and {\tt mvp}
packages~\cite{hankin2022_mvp,hankin2022_spray}.  Clifford objects are
thus considered to be the sum of a finite number of {\bf blades}, each
multiplied by a coefficient.  One reason why the {\tt map} class is
fast is that the order in which the keys are stored is undefined: the
compiler may store them in the order which it regards as most
propitious.  This is not an issue for the maps considered here as
addition and multiplication are commutative and associative.  The
package uses {\tt disordR} discipline.

The terms of a {\tt clifford} object are held in an
implementation-specific order and dealing with this in a consistent
way is achieved using the {\tt disordR}
package~\cite{hankin2022_disordR}, which presents an extended
discussion.

\section{The package in use}

Suppose we want to work with Clifford object
$1+2\ei{1}+3\ei{2}+4\ei{2}\ei{3}$.  In R idiom this would be

\begin{Schunk}
\begin{Sinput}
> (x <- 1 + 2*e(1) + 3*e(2) + 4*e(2:3))
\end{Sinput}
\begin{Soutput}
Element of a Clifford algebra, equal to
+ 1 + 2e_1 + 3e_2 + 4e_23
\end{Soutput}
\end{Schunk}

Here we have used function {\tt e()} which takes an integer vector that
specifies the term.  Addition and subtraction work as expected:

\begin{Schunk}
\begin{Sinput}
> y <- e(1) + 55*e(1:5)
> x-y
\end{Sinput}
\begin{Soutput}
Element of a Clifford algebra, equal to
+ 1 + 1e_1 + 3e_2 + 4e_23 - 55e_12345
\end{Soutput}
\end{Schunk}

In the above, see how the $\ei{1}$ term has vanished.  We can
multiply Clifford elements using natural R idiom:

\begin{Schunk}
\begin{Sinput}
> x*x
\end{Sinput}
\begin{Soutput}
Element of a Clifford algebra, equal to
- 2 + 4e_1 + 6e_2 + 8e_23 + 16e_123
\end{Soutput}
\end{Schunk}

(Multiplication that Snygg denotes by juxtaposition is here indicated
with a {\tt *}).  We can consider arbitrarily high dimensional data:

\begin{Schunk}
\begin{Sinput}
> (z <- as.1vector(1:7))
\end{Sinput}
\begin{Soutput}
Element of a Clifford algebra, equal to
+ 1e_1 + 2e_2 + 3e_3 + 4e_4 + 5e_5 + 6e_6 + 7e_7
\end{Soutput}
\begin{Sinput}
> z*x
\end{Sinput}
\begin{Soutput}
Element of a Clifford algebra, equal to
+ 8 + 1e_1 - 10e_2 - 1e_12 + 11e_3 - 6e_13 - 9e_23 + 4e_123 + 4e_4 - 8e_14 -
12e_24 + 16e_234 + 5e_5 - 10e_15 - 15e_25 + 20e_235 + 6e_6 - 12e_16 - 18e_26 +
24e_236 + 7e_7 - 14e_17 - 21e_27 + 28e_237
\end{Soutput}
\end{Schunk}

In the above, we coerce a vector to a Clifford 1-vector.  The package
includes many functions to generate Clifford objects:

\begin{Schunk}
\begin{Sinput}
> rcliff()
\end{Sinput}
\begin{Soutput}
Element of a Clifford algebra, equal to
+ 4 + 1e_1 + 5e_2 + 4e_13 - 2e_4 - 3e_135 + 2e_1345 + 3e_36 - 1e_56
\end{Soutput}
\end{Schunk}

The defaults for {\tt rcliff()} specify that the object is a sum of
grade-4 terms but this can be altered:

\begin{Schunk}
\begin{Sinput}
> (x <- rcliff(d=7,g=5,include.fewer=TRUE))
\end{Sinput}
\begin{Soutput}
Element of a Clifford algebra, equal to
+ 4 - 2e_2 + 1e_23 + 3e_14 + 2e_6 + 5e_237 - 3e_257 + 4e_1357 - 1e_67
\end{Soutput}
\begin{Sinput}
> grades(x)
\end{Sinput}
\begin{Soutput}
A disord object with hash 3ae4a22758225c262361edbf478a16c0b1551f50 and elements
[1] 0 1 2 2 1 3 3 4 2
(in some order)
\end{Soutput}
\end{Schunk}

\section{Pseudo-Euclidean spaces}

The signature of the metric may be altered.  Starting with the
Euclidean case we have:
    
\begin{Schunk}
\begin{Sinput}
> e1 <- e(1)
> e2 <- e(2)
> e1*e1
\end{Sinput}
\begin{Soutput}
Element of a Clifford algebra, equal to
scalar ( 1 )
\end{Soutput}
\begin{Sinput}
> e2*e2
\end{Sinput}
\begin{Soutput}
Element of a Clifford algebra, equal to
scalar ( 1 )
\end{Soutput}
\end{Schunk}

(function {\tt e(i)} returns $\ei{i}$).  However, if we wish to
consider $n=\begin{bmatrix}1&0\\0&-1\end{bmatrix}$, the package idiom
is to use the {\tt signature()} function:

\begin{Schunk}
\begin{Sinput}
> signature(1,1)  # signature +-
> e1*e1 # as before, returns +1
\end{Sinput}
\begin{Soutput}
Element of a Clifford algebra, equal to
scalar ( 1 )
\end{Soutput}
\begin{Sinput}
> e2*e2 # should return -1
\end{Sinput}
\begin{Soutput}
Element of a Clifford algebra, equal to
scalar ( -1 )
\end{Soutput}
\end{Schunk}

Suppose we wish to use a signature $+++-$, corresponding to the
Minkowski metric in special relativity; this would be indicated in
package idiom by {\tt signature(3,1)}.  Note that the clifford objects
themselves do not store the signature; it is used only by the product
operation {\tt *}.

\begin{Schunk}
\begin{Sinput}
> x <- rcliff(d=4,g=3,include.fewer=TRUE)
> y <- rcliff(d=4,g=3,include.fewer=TRUE)
\end{Sinput}
\end{Schunk}

Then we may multiply these two clifford objects using either the
default positive-definite inner product, or the Minkowski metric:

\begin{Schunk}
\begin{Sinput}
> x*y
\end{Sinput}
\begin{Soutput}
Element of a Clifford algebra, equal to
+ 16 + 4e_1 + 4e_2 + 1e_12 - 16e_3 - 1e_13 + 1e_23 - 8e_123 + 16e_4 + 4e_14 +
12e_24 + 3e_124 - 2e_34 + 8e_134 + 48e_234 + 5e_1234
\end{Soutput}
\begin{Sinput}
> signature(3,1)  # switch to signature +++-
> x*y
\end{Sinput}
\begin{Soutput}
Element of a Clifford algebra, equal to
+ 39 + 10e_1 + 19e_2 - 1e_12 - 40e_3 - 1e_13 - 24e_23 - 8e_123 + 19e_4 + 2e_14
+ 31e_24 + 9e_124 + 6e_34 + 8e_134 + 48e_234 + 5e_1234
\end{Soutput}
\end{Schunk}

In the above, see how the products are different using the two inner
products.  

\section{Grassman algebra}

A Grassman algebra corresponds to a Clifford algebra with identically
zero inner product.  Package idiom is to use a zero signature:

\begin{Schunk}
\begin{Sinput}
> signature(0,0)  # specify null inner product
\end{Sinput}
\end{Schunk}
\begin{Schunk}
\begin{Sinput}
> is.zero(e(5)^2)     # should be TRUE
\end{Sinput}
\begin{Soutput}
[1] TRUE
\end{Soutput}
\end{Schunk}

This is a somewhat clunky way of reproducing the functionality of the
{\tt stokes} package~\cite{hankin2022_stokes,hankin2022_stokes_arxiv}.
If we have

\begin{Schunk}
\begin{Sinput}
> x <- clifford(list(1:3, c(2,3,7)), coeffs=3:4)
> y <- clifford(list(1:3, c(1,4,5), c(4,5,6)), coeffs=1:3)
> x %^% y
\end{Sinput}
\begin{Soutput}
Element of a Clifford algebra, equal to
+ 9e_123456 - 8e_123457 - 12e_234567
\end{Soutput}
\end{Schunk}

then the {\tt stokes} idiom for this would be:

\begin{Schunk}
\begin{Sinput}
> (x <- as.kform(rbind(1:3,c(2,3,7)),3:4))
\end{Sinput}
\begin{Soutput}
           val
 2 3 7  =    4
 1 2 3  =    3
\end{Soutput}
\begin{Sinput}
> (y <- as.kform(rbind(1:3,c(1,4,5),4:6),1:3))
\end{Sinput}
\begin{Soutput}
           val
 1 2 3  =    1
 1 4 5  =    2
 4 5 6  =    3
\end{Soutput}
\begin{Sinput}
> x %^% y
\end{Sinput}
\begin{Soutput}
                 val
 1 2 3 4 5 6  =    9
 2 3 4 5 6 7  =  -12
 1 2 3 4 5 7  =   -8
\end{Soutput}
\end{Schunk}

\section{Positive-definite inner product}

Function {\tt signature()} takes an infinite argument to make the
inner product positive-definite:

\begin{Schunk}
\begin{Sinput}
> signature(Inf)
\end{Sinput}
\end{Schunk}

(internally the package sets the signature to
{\tt .Machine\$integer.max}, a near-infinite integer).  With this,
$\ei{i}\ei{i}=+1$ for any $i$:

\begin{Schunk}
\begin{Sinput}
> e(53)^2
\end{Sinput}
\begin{Soutput}
Element of a Clifford algebra, equal to
scalar ( 1 )
\end{Soutput}
\end{Schunk}

\section{Left and right contractions}

\cite{dorst2002} defines the left contraction $A\rfloor B$ and right
contraction $A\lfloor B$ (\cite{chisholm2012} calls these left and
right inner products) as follows:

\begin{eqnarray}
\displaystyle A\rfloor B = \sum_{r,s}\left\langle\left\langle
A\right\rangle_r\left\langle B\right\rangle_s\right\rangle_{s-r}\\
\displaystyle A\lfloor B = \sum_{r,s}\left\langle\left\langle
A\right\rangle_r\left\langle B\right\rangle_s\right\rangle_{r-s}
\end{eqnarray}

Package idiom for these would be {\tt A\%\_|\%B} and {\tt A\%|\_\%B}
---or {\tt lefttick(A,B)} and {\tt righttick(A,B)}---respectively.
Thus:

\begin{Schunk}
\begin{Sinput}
> (A <- rcliff())
\end{Sinput}
\begin{Soutput}
Element of a Clifford algebra, equal to
+ 4 + 1e_3 - 3e_23 - 2e_1234 - 1e_1236 + 4e_2346 + 3e_156 + 5e_256 + 2e_2356
\end{Soutput}
\begin{Sinput}
> (B <- rcliff())
\end{Sinput}
\begin{Soutput}
Element of a Clifford algebra, equal to
+ 4 + 2e_134 + 3e_125 + 5e_1235 + 4e_245 - 2e_1245 - 1e_346 - 3e_356 + 1e_1456
\end{Soutput}
\begin{Sinput}
> A %_|% B
\end{Sinput}
\begin{Soutput}
Element of a Clifford algebra, equal to
+ 16 - 3e_4 - 2e_14 + 8e_134 + 15e_15 + 17e_125 + 20e_1235 + 16e_245 - 8e_1245
- 1e_46 - 4e_346 - 3e_56 - 12e_356 + 4e_1456
\end{Soutput}
\begin{Sinput}
> A %|_% B
\end{Sinput}
\begin{Soutput}
Element of a Clifford algebra, equal to
+ 16 + 6e_2 + 4e_3 - 12e_23 - 8e_1234 - 4e_1236 + 16e_2346 + 12e_156 + 20e_256
+ 8e_2356
\end{Soutput}
\end{Schunk}

One thing to be wary of is the order of operations.  Thus
$\ei{2}\rfloor\ei{12}=-\ei{1}$ (in a positive-definite space) but

\begin{Schunk}
\begin{Sinput}
> e(2) %_|% e(1)*e(2)
\end{Sinput}
\begin{Soutput}
Element of a Clifford algebra, equal to
the zero clifford element (0)
\end{Soutput}
\end{Schunk}

because this is parsed as $(\ei{2}\rfloor\ei{1})\ei{2}=0\ei{2}=0$.  To
evaluate this as intended we need to include brackets:

\begin{Schunk}
\begin{Sinput}
> e(2) %_|% (e(1)*e(2))
\end{Sinput}
\begin{Soutput}
Element of a Clifford algebra, equal to
- 1e_1
\end{Soutput}
\end{Schunk}

although in this case it might be preferable to create the terms directly:

\begin{Schunk}
\begin{Sinput}
> e(2) %_|% e(1:2)
\end{Sinput}
\begin{Soutput}
Element of a Clifford algebra, equal to
- 1e_1
\end{Soutput}
\end{Schunk}

\subsection{Numerical verification of left and right inner product identities}

Chisholm gives a number of identities for these products including

\begin{eqnarray}
  A\rfloor(B\lfloor C) &=& (A\rfloor B)\lfloor C\\
  A\rfloor(B\rfloor C) &=& (A\wedge B)\rfloor C\\
  A\lfloor(B\wedge  C) &=& (A\lfloor B)\lfloor C
\end{eqnarray}

In package idiom:

\begin{Schunk}
\begin{Sinput}
> A <- rcliff();  B <- rcliff();  C <- rcliff()
> A %_|% (B %|_% C) == (A %_|% B) %|_% C
\end{Sinput}
\begin{Soutput}
[1] TRUE
\end{Soutput}
\begin{Sinput}
> A %_|% (B %_|% C) == (A %^%  B) %_|% C
\end{Sinput}
\begin{Soutput}
[1] TRUE
\end{Soutput}
\begin{Sinput}
> A %|_% (B %^%  C) == (A %|_% B) %|_% C
\end{Sinput}
\begin{Soutput}
[1] TRUE
\end{Soutput}
\end{Schunk}
\section{Higher dimensional spaces}

Ab\l{}amowicz and Fauser~\cite{ablamowicz2012} consider
high-dimensional Clifford algebras and consider the following two
elements of the 1024-dimensional Clifford algebra which we may treat
as ${\mathcal C}_{7,3}$ spanned by $\ei{1},\ldots,\ei{10}$ and perform
a calculation which I reproduce below (although these authors
exploited Bott periodicity, a feature not considered here).

Firstly we change the default print method slightly:

\begin{Schunk}
\begin{Sinput}
> options("basissep" = ",")
\end{Sinput}
\end{Schunk}

(this separates the subscripts of the basis vectors with a comma,
which is useful for clarity if $n>9$).  We then define clifford
elements $x,y$:

\begin{Schunk}
\begin{Sinput}
> (x <- clifford(list(1:3,c(1,5,7,8,10)),c(4,-10)) + 2)
\end{Sinput}
\begin{Soutput}
Element of a Clifford algebra, equal to
+ 2 + 4e_1,2,3 - 10e_1,5,7,8,10
\end{Soutput}
\begin{Sinput}
> (y <- clifford(list(c(1,2,3,7),c(1,5,6,8),c(1,4,6,7)),c(4,1,-3)) - 1)
\end{Sinput}
\begin{Soutput}
Element of a Clifford algebra, equal to
- 1 + 4e_1,2,3,7 - 3e_1,4,6,7 + 1e_1,5,6,8
\end{Soutput}
\end{Schunk}

Their geometric product is given in the package as

\begin{Schunk}
\begin{Sinput}
> signature(7)
> x*y
\end{Sinput}
\begin{Soutput}
Element of a Clifford algebra, equal to
- 2 - 4e_1,2,3 - 16e_7 + 8e_1,2,3,7 - 6e_1,4,6,7 - 12e_2,3,4,6,7 + 2e_1,5,6,8 +
4e_2,3,5,6,8 - 40e_2,3,5,8,10 - 30e_4,5,6,8,10 + 10e_1,5,7,8,10
\end{Soutput}
\end{Schunk}

in agreement with \cite{ablamowicz2012}, although the terms appear in
a different order.

\section{Conclusions and further work}

The {\tt clifford} package furnishes a consistent and documented suite
of reasonably efficient {\tt R}-centric functionality.  Further work
might include closer integration with the {\tt stokes}
package~\citep{hankin2022_stokes,hankin2022_stokes_arxiv}.

\bibliographystyle{plain}
\bibliography{clifford_arxiv}

\begin{thebibliography}{10}

\bibitem{ablamowicz2012}
Rafal Ab\l{}amowicz and Bertfried Fauser.
\newblock Symbolic computations in higher dimensional clifford algebras, 2012.

\bibitem{chisholm2012}
Eric Chisolm.
\newblock Geometric algebra, 2012.

\bibitem{dorst2002}
L.~Dorst.
\newblock {\em Applications of geometric algebra in computer science and
  engineering}, chapter~2, pages 35--46.
\newblock Birkh\"{a}user, 2002.

\bibitem{hankin2022_disordR}
Robin K.~S. Hankin.
\newblock Disordered vectors in {R}: introducing the {\tt disordr} package,
  2022.

\bibitem{hankin2022_mvp}
Robin K.~S. Hankin.
\newblock Fast multivariate polynomials in {R}: the {{\tt mvp}} package.
\newblock \url{https://arxiv.org/abs/2210.15991}, 2022.

\bibitem{hankin2022_spray}
Robin K.~S. Hankin.
\newblock Sparse arrays in {R}: the {{\tt spray}} package.
\newblock \url{https://arxiv.org/abs/2210.03856}, 2022.

\bibitem{hankin2022_stokes}
Robin K.~S. Hankin.
\newblock {\em {\tt stokes}: the exterior calculus}, 2022.
\newblock {R} package version 1.1-6.

\bibitem{hankin2022_stokes_arxiv}
Robin K.~S. Hankin.
\newblock Stokes's theorem in {R}.
\newblock \url{https://arxiv.org/abs/2210.17008}, 2022.

\bibitem{hestenes1987}
D.~Hestenes and G.~Sobczyk.
\newblock {\em Clifford algebra to geometric calculus}.
\newblock Kluwer, 1987.

\bibitem{sympy2017}
Aaron Meurer et~al.
\newblock Sympy: symbolic computing in python.
\newblock {\em PeerJ Computer Science}, 3:e103, January 2017.

\bibitem{musser2009}
David~R. Musser, Gillmer~J. Derge, and Atul Saini.
\newblock {\em {STL} Tutorial and Reference Guide: {C++} Programming with the
  Standard Template Library}.
\newblock Addison-Wesley Professional, 3rd edition, 2009.

\bibitem{rcore2022}
{R Core Team}.
\newblock {\em R: A Language and Environment for Statistical Computing}.
\newblock R Foundation for Statistical Computing, Vienna, Austria, 2022.

\bibitem{snygg2010}
J.~Snygg.
\newblock {\em A new approach to differential geometry using {C}lifford's
  geometric algebra}.
\newblock Birkh\"{a}user, 2010.

\bibitem{sagemath2019}
{The Sage Developers}.
\newblock {\em {S}ageMath, the {S}age {M}athematics {S}oftware {S}ystem
  ({V}ersion 8.6)}, 2019.

\end{thebibliography}

\end{document}